\documentclass[a4paper, 12pt]{article}
\author{C. J. de Matos\footnote{ESA-HQ, European Space Agency, 8-10 rue Mario Nikis, 75015 Paris, France, e-mail: Clovis.de.Matos@esa.int}
}
\title{Gravitomagnetic London Moment in Rotating Supersolid $He^4$}

\begin{document}

\maketitle \begin{abstract}Non classical rotational inertia
observed in rotating supersolid $He^4$ can be accounted for by a
gravitomagnetic London moment similar to the one recently reported
in rotating superconductive rings.
\end{abstract}

Non Classical Rotational Inertia (NCRI) was predicted by London 50
years ago \cite{London}. It was eventually verified experimentally
by Hess and Fairbank \cite{Hess}, who set Helium in a suspended
bucket into rotation above the critical temperature $T_c$ at which
superfluidity sets in, and then cooled it (with the bucket still
rotating at angular velocity $\omega$) through $T_c$. They found
that, provided $\omega$ is less than a critical value $\omega_c$,
the apparent moment of inertia of the helium - that is, the ratio
of its angular momentum to $\omega$ - is not given by the
classical value $I_{classical}$ but rather by
\begin{equation}
I(T)=\frac{L}{\omega}=I_{classical}\Bigg[1-f_s(T)\Bigg]\label{1}
\end{equation}
where $f_s(T)=\rho^*/\rho$ is the superfluid fraction. In liquid
helium, $f_s(T)$ tends to 1 in the zero-temperature limit and to
zero when $T=T_c$, $\rho^*$ and $\rho$ being respectively the mass
density of superfluid helium, and the mass density of normal
helium.

\begin{figure}[!h]
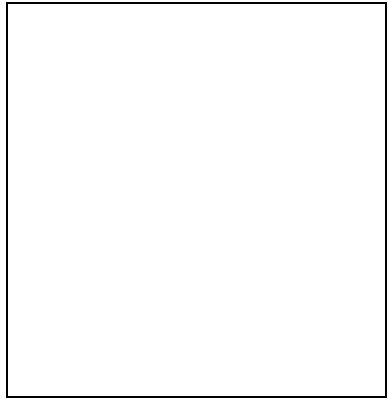

\makebox[\textwidth]{\framebox[5cm]{\rule{0pt}{5cm}}}
\caption{Signs of supersolidity. The supersolid fraction $f_s(T)$
inferred from the data in \cite{Kim} as interpreted by
Eq.(\ref{1}), as a function of temperature for different values of
the maximum velocity of the walls. The pressure is $41 \, Bars$.
[Adapted from \cite{Kim}] \label{chan}}
\end{figure}

NCRI has recently been observed in rotating supersolid $He^4$ by
Kim and Chan \cite{Kim}. They measured the resonance frequency of
a torsional oscillator that contains an annulus of solid $He^4$.
Below $230 mK$, the frequency experiences a relative increase that
depends on the temperature and drive amplitude and reaches a
maximum of about four parts in $10^5$. Having excluded by various
control experiments, other explanations, they conclude that the
data indicate a change in the moment of inertia of the supersolid,
which according to Equ.(\ref{1}), corresponds to a maximum
supersolid fraction ($f_s(T)=\rho^*_{supersolid \,
He^4}/\rho_{normal \, solid \, He^4}$) of $~0.017$ (see
figure~\ref{chan}).

Tajmar and the author recently \cite{Tajmar1} \cite{de Matos1}
\cite{Tajmar2} observed a gravitomagnetic London moment, $B_g \,
[Rad/s]$, in rotating superconductive rings.
\begin{equation}
B_g=2\omega f_s(T)\label{2}
\end{equation}
Where $\omega$ is the angular velocity of the ring, and
$f_s(T)=\rho^*/\rho$ is the Cooper pairs fraction, $\rho^*$ being
the Cooper pairs mass density and $\rho$ the superconductor's bulk
density.

Assuming that a rotating supersolid $He^4$ crystal also exhibits a
gravitomagnetic London moment, $B_g$ proportional to the
supersolid fraction, its angular momentum would be given by
\begin{equation}
L=I_{classical}\Bigg[\omega-\frac{1}{2} B_g\Bigg]\label{3}
\end{equation}
Doing Equ.(\ref{2}) into Equ.(\ref{3}) we find back Equ.(\ref{1})!

Therefore we conclude that the rotation of supersolid $He^4$
exhibits a gravitomagnetic London moment similar to the one
observed in rotating superconductive rings, which can account for
the observed NCRI in this physical system. Kim and Chan's
experiment would tend to confirm the existence of the
gravitomagnetic London moment in rotating quantum materials.

\end{document}